\documentstyle[prl,aps,epsf,twocolumn]{revtex}

\title{Atomic Scattering in Presence of an External Confinement \\
and a Gas of Impenetrable Bosons}

\author
{
M. Olshanii
}
\address
{
Lyman Laboratory, Harvard University,
Cambridge, MA 02138, USA
\\
{\small and}
\\
Ecole Normale Sup\'{e}rieure,
Laboratoire Kastler-Brossel,
24 Rue Lhomond,
75231 Paris Cedex 05, France \\
{\small E-mail: {\it maxim@atomsun.harvard.edu}}
}

\date{January 22, 1998; accepted in PRL June 22, 1998}

\begin{document}
\maketitle
\begin{abstract}
We calculate, within the pseudopotential approximation,  
a one-dimensional scattering amplitude and effective 
one-dimensional interaction potential for atoms confined transversally      
by an atom waveguide or highly elongated ``cigar''-shaped atomic trap. 
We show that in the low-energy scattering regime, the scattering process 
degenerates 
to a total reflection suggesting an experimental 
realization of a famous model in theoretical physics - a one-dimensional 
gas of impenetrable 
bosons (``Tonks'' gas). We give an estimate for suitable experimental 
parameters for 
alkali atoms confined in waveguides. 
\end{abstract}
%
%



Rapid progress in producing
Bose condensates of alkali atoms \cite{Eric-Wolfgang-Hulett}
opened up new areas of ultra-low energy collisional physics.
Concurrently, work has progressed on confinement of atoms in the 
light-induced \cite{Waveguides} and 
magnetic-field-induced \cite{Magnetic_Guides} atom waveguides. 

We develop a theory for the binary atomic collisions in presence of 
a transverse external confinement. 
Using the pseudopotential approximation, we derive
an expression for an effective one-dimensional scattering amplitude
and show that the interparticle interaction can be approximated  
by an effective one-dimensional $\delta$-potential of a known strength.
In the case of a dilute atomic gas, when the three body 
collisions are negligible, our effective potential can also be  used  
to describe quasi-one-dimensional many-body systems:
a project on the experimental realization of  the one-dimensional 
Bose condensate 
is already presented in the literature \cite{1D_BEC}.
Results of our paper will allow one to properly take into
account the trap-induced corrections to 
the strength of the atomic mean-field potential.

Furthermore, we show that in the low-energy scattering 
limit ($k_{z} |a_{\rm 1D}| \ll 1$), the effective one-dimensional scattering 
degenerates to
a total reflection ($\hbar k_{z}$ is the longitudinal component of the 
atomic momentum 
and $a_{\rm 1D}$ is the
one-dimensional scattering length defined below). This conclusion 
allows us to suggest
an experimental realization of another famous
model in theoretical physics -- a one-dimensional gas of impenetrable bosons
\cite{Imp_Bosons} also referred to as a ``Tonks'' gas. Such a system provides 
us with an unusual example of
a boson-fermion duality: the elementary excitations of such a bosonic system 
obey Fermi statistics \cite{Thermodynamics_of_Imp_Bosons} 
(which is possible only in one dimension \cite{Korepin}).
The Tonks gas is, therefore, 
a system complementary to the one-dimensional Bose condensate:
the latter requires the high-energy scattering regime
($k_{z} |a_{\rm 1D}| \gg 1$) \cite{Imp_Bosons_Bogoliubov}
and
its excitation spectrum is represented
by the Bogoliubov bosons.

%
To describe the binary collisions between cold atoms confined in  
a waveguide, we suggest the following model:
\begin{description}
\item (a) The waveguide potential is replaced by an axially symmetric 
2D harmonic potential
of a frequency $\omega_{\perp}$. The forces created by the potential 
act along the X-Y plane;
\item (b) Atomic motion along the Z axis is free; 
\item
(c) Interaction between the atoms is modeled by the Huang's  pseudopotential
\cite{Huang}
\begin{eqnarray}
U(r) =
g \delta({\bf r}) \, \frac{\partial}{\partial r} (r \, \cdot \,) \, \, ,
\label{Huang}
\end{eqnarray}
where
$g=2\pi\hbar^2 a/\mu$ is the potential strength, $a$ is the
$s$-wave scattering length for the ``true'' interaction potential,
$\mu = m/2$ is the reduced mass, $m$ is the mass of the atoms. 
The regularization operator 
$\frac{\partial}{\partial r} (r \, \cdot \,)$, that removes 
the $1/r$ divergence from the scattered wave (see \cite{Huang}), plays 
an important role in the derivation below;
\item (d) Atomic motion (both transverse and longitudinal
components) is ``cooled down'' below 
the transverse vibrational energy
$\hbar \omega_{\perp}$. We will justify this condition
in the second subsequent paragraph.
\end{description}

The harmonic nature of the confining potential 
allows the separation of
the center-of-mass and relative motions.
The Schr\"{o}dinger equation governing the relative motion reads
\begin{eqnarray}
\left\{
 \frac{\hat{p}_z^2}{2\mu}
 +g \delta({\bf r}) \, \frac{\partial}{\partial r} (r \, \cdot \,)  
 + \hat{H}_{\perp}
   (\hat{p}_x, \hat{p}_y, x, y)  
\right\} \Psi
= E \Psi \, ,
\label{C_Schrodinger}
\end{eqnarray}
where ${\bf r} = {\bf r_2} - {\bf r_1}$ is a relative 
coordinate for atoms $1$ and $2$ and
\begin{eqnarray}
\hat{H}_{\perp}
= \frac{\hat{p}_x^2+\hat{p}_y^2}{2\mu}
   + \frac{\mu\omega_{\perp}^2 (x^2+y^2)}{2}
\label{C_Trans}
\end{eqnarray}
is a 2D harmonic oscillator Hamiltonian. 

We will suppose that 
(i) the incident wave 
$e^{ik_{z}z}\phi_{n=0,m_{z}=0}(\rho)$ corresponds 
to a particle in the ground state of the transverse 
Hamiltonian (\ref{C_Trans}) ($\rho = \sqrt{x^2+y^2}$), 
and 
(ii) the longitudinal kinetic energy of the incident wave
is limited by the energy spacing  between the ground and  
first {\it axially symmetric} excited state: 
\begin{eqnarray}
\frac{\hbar^2 k_{z}^2}{2\mu} < E_{n=2, m_{z}=0} - E_{n=0,m_{z}=0} 
= 2\hbar\omega_{\perp}\, . 
\label{C_Kinetic_Constr}
\end{eqnarray}
Here 
$E_{n, m_{z}} = \hbar\omega_{\perp}(n+1)$ is the energy spectrum of the 2D harmonic 
oscillator, $n=0,1,2,\ldots,\infty$ is the principal quantum number, 
$m_{z}$ is the angular momentum
with respect to the Z axis, $m_{z} = 0,2,4,\ldots,n \, (1,3,5,\ldots,n)$ 
if $n$ is even(odd). 
The asymptotic form of the scattering wave function $\Psi$
then reads
\begin{eqnarray}
&\hspace{-5.5cm}\Psi(z,\rho)
\stackrel{|z| \rightarrow \infty}{\longrightarrow}
\nonumber \\
&\left\{ \,
e^{ik_{z}z}
+ f_{\rm even} \, e^{ik_{z}|z|} 
+ f_{\rm odd} \, {\rm sign}(z)e^{ik_{z}|z|}
\, \right\} \, \phi_{0,0}(\rho) \, ,
\label{Psi_Asymp}
\end{eqnarray}
where 
the first term in the 
curly brackets represents the incident wave, the second and third 
terms give the even and odd scattered waves respectively, and
$f_{\rm even}(k_{z})$ and $f_{\rm odd}(k_{z})$  are  
the {\it one-dimensional scattering amplitudes} 
for the even and odd partial waves respectively. 
Note that the transverse state ($n=0, \, m_{z}=0$) remains unchanged 
after the collision.
Transitions to the higher ($n>0$) modes are forbidden due to
either angular momentum or energy conservation 
(if the
condition (\ref{C_Kinetic_Constr}) is satisfied).
We would like to stress however, that 
during the collision itself a virtual
excitation of the high energy axially symmetric modes ($n>0, \, m_{z}=0$) 
is not forbidden and it is properly taken into account below. 
One of the goals of this paper is to calculate  
the amplitudes (\ref{Psi_Asymp}).  

For the zero-range potential (\ref{Huang}) 
the one-dimensional scattering amplitudes (\ref{Psi_Asymp}) 
can be calculated analytically. To perform this calculation 
we expand the wave function $\Psi(z,\rho)$ to a series
over the eigenstates of the transverse Hamiltonian (\ref{C_Trans}),
substitute the expansion into the Schr\"{o}dinger 
equation (\ref{C_Schrodinger}) using 
$E = \hbar^2 k_{z}^2/2\mu + \hbar\omega_{\perp}$ for the energy 
in the right hand side, and then apply the 
asymptotic conditions (\ref{Psi_Asymp}) along with  
the conditions 
of the continuity of the wave function and its derivative. 
We obtain the following expression for the 
scattering amplitudes:
\begin{eqnarray}
f_{\rm even} = 
-\, \frac{i \mu}{\hbar^2 k_{z}} \phi_{0,0}^*(0)g\eta 
\,\,\,\,\,\, ; \,\,\,\,\,\, 
f_{\rm odd} = 0 \,\, ,
\end{eqnarray}
where $\eta$ is the $r \rightarrow 0$ limit of the 
regular (free of the $1/r$ divergence) part 
of the solution $\Psi$: 
\begin{eqnarray}
\eta =  
\frac{\partial}{\partial r} 
\lbrack r \, \Psi({\bf r}) \rbrack \Big|_{r \rightarrow 0}
= 
\frac{\partial}{\partial z} \lbrack z \, \Psi(z, \rho=0)\rbrack
\Big|_{z \rightarrow 0+}
\,\,\,\, .
\label{Eta_Equation}
\end{eqnarray}
The expression for the wave function reads
\begin{eqnarray}
\Psi(z, \rho=0) &=& 
\frac{1}{\sqrt{\pi} a_{\perp}} \exp(i k_{z} z ) 
\nonumber \\
&-& \frac{i g \mu \eta}{\pi \hbar^2 k_{z} a_{\perp}^2} 
                 \exp(i k_{z} |z| )
\nonumber \\
&-& \frac{ g \mu \eta}{2 \pi \hbar^2 a_{\perp}} 
                 \Lambda
                   \left\lbrack
                     \frac{2|z|}{a_{\perp}}, \, 
                     -\left(\frac{k_{z}a_{\perp}}{2}\right)^2
                   \right\rbrack \, ,
\end{eqnarray}
where the function $\Lambda$ is defined as 
$
\Lambda\lbrack \xi, \, \epsilon \rbrack =
\sum_{s^{\prime}=1}^{\infty}
\exp(-\sqrt{s^{\prime}+\epsilon}\,\xi)/\sqrt{s^{\prime}+\epsilon}
$ \cite{Epsilon}, 
the sum over $s^{\prime}=n/2$ originates from a sum 
$\sum_{n=0,2,4,\ldots}^{\infty}\ldots \,\,$
over the transverse states $\phi_{n,m_{z}=0}$,
and $a_{\perp} = (\hbar/\mu\omega_{\perp})^{1/2}$ is the size
of the ground state $\phi_{n=0,m_{z}=0}$ of transverse Hamiltonian (\ref{C_Trans}).
Above we have used a simple relation 
$|\phi_{n,m_{z}=0}(\rho=0)|^2 =
1/\pi a_{\perp}^2$ for the wave functions of the 2D harmonic oscillator. 
The value $\phi_{0,0}(\rho=0)$
has been chosen to be real and positive. 

The regular part $\eta$ of the wave function is not explicitly 
defined yet, being a solution of the equation (\ref{Eta_Equation}).
Note that the order of the partial derivative  
$\partial/\partial z$ and the sum over $s^{\prime}$ in the $\Lambda$-function 
can not be interchanged because the series $\Lambda$ does not
converge uniformly as $\xi \rightarrow 0$. The value of $\eta$  
can be extracted from the equation (\ref{Eta_Equation}) using   
the following expansion for the function $\Lambda$:  
$\Lambda\lbrack \xi, \, \epsilon \rbrack 
= 2/\xi + {\cal L}(\epsilon) + {\cal L}_1(\epsilon)\xi +\ldots \,\,$, 
where the zero-order term of the expansion has a form 
${\cal L}(\epsilon) = - \,{\cal C} + \bar{\cal L}(\epsilon)$ and
\begin{eqnarray}
&{\cal C}&=\lim_{s \rightarrow \infty} \,
\left(
\int_{0}^{s} \frac{d\, s^{\prime}}{\sqrt{s^{\prime}}}
-\sum_{s^{\prime}=1}^{s} \,
\frac{1} {\sqrt{s^{\prime}}}
\right)
= 1.4603\ldots \,\,\, , 
\label{C_Constant}\\
&\bar{\cal L}&(\epsilon) = \sum_{n=1}^{\infty} (-1)^n \,
\frac
 {
  {\zeta}[(1 + 2n)/2] (2n-1)!! \, \epsilon^n
 }
 {
  2^n n!
 } \, \, . 
\label{Q_Bar_Function}
\end{eqnarray}
To prove this formula 
one should subtract and add a sum
$
\sum_{s^{\prime}=1}^{\infty}
\int_{s^{\prime}-1}^{s^{\prime}} \, ds^{\prime\prime} \,
\exp(-\sqrt{s^{\prime\prime}}\xi)/\sqrt{s^{\prime\prime}}
=
2/\xi
\label{2/xi}
$
to the function $\Lambda$. The sum of the differences between the $\Lambda$
terms and the terms of the above sum 
converges uniformly and the derivative in (\ref{Eta_Equation})
can be easily calculated. 
Here ${\zeta}[\xi]$ is the Riemann zeta-function.

We now write down the final expression for the one-dimensional scattering 
amplitude (\ref{Psi_Asymp}): 
\begin{eqnarray}
f_{\rm even}(k_{z})  = & & 
\nonumber\\
& &\hspace{-1cm} - \, \frac{1}
{
 1 + i k_{z} a_{\rm 1D}
 -
 \underbrace
   {
    (i k_{z} a_{\perp}/2) \,\, \bar{\cal L} 
         ( - k_{z}^2 a_{\perp}^2 / 4 )
   }_{{\cal O}((k_{z}a_{\perp})^3)} 
} \,\,\,\,\,\, ,
\label{C_Amplitude}
\end{eqnarray}
where the function $\bar{\cal L}(\epsilon)$ is given by (\ref{Q_Bar_Function}).
Here 
\begin{eqnarray}
a_{\rm 1D} = - \frac{a_{\perp}^2}{2 a} \,
\left(
 1-{\cal C}\frac{a}{a_{\perp}}
\right)
\label{a_1D} 
\end{eqnarray}
is  
the {\it one-dimensional scattering length} and the constant ${\cal C}$
is given by (\ref{C_Constant}). In analogy with three-dimensional
scattering, the one-dimensional scattering length is defined as 
a derivative $-\partial\Delta/\partial k_{z}|_{k_{z}\rightarrow 0+}$ 
of the even-wave scattering phase. The scattering phase 
$\Delta(k_{z})$ is defined through   
the even solution 
$\Psi \propto \sin\lbrack k_{z}|z| + \Delta(k_{z}) \rbrack \phi_{0,0}(\rho)$
of the Schr\"{o}dinger equation (\ref{C_Schrodinger}). 
The  
formula for the one-dimensional
scattering amplitude (\ref{C_Amplitude}) is the key result of this paper.

The expression (\ref{C_Amplitude}) is valid for any strength  
of the transverse confinement. Note however, that 
the tight confinement limit $a_{\perp} \ll |a|$ makes sense only
if the $s$-wave scattering amplitude for 
the ``true'' 3D finite-range interatomic potential (approximated 
in this paper by
the Huang's pseudopotential (\ref{Huang})) shows a
zero-energy resonance;
namely if the $s$-wave scattering length $a$ 
(the same for both potentials) is much greater than
the effective range $r_0$  of the ``true'' potential \cite{Resonant_Scatt}.
Indeed the 3D pseudopotential
approximation (\ref{Huang}) is only valid for 
velocities which are lower then the inverted effective range
($k \ll r_0^{-1}$) (see \cite{Landau}), and therefore requires
a lower bound for the
transverse size: $a_{\perp} \sim k_{\perp}^{-1} \gg r_0$.
This  condition and the tight confinement regime
criterion ($a_{\perp} \ll |a|$) are consistent only
in the resonant
case: $|a| \gg r_0$.

%
%
%
\begin{figure}
\begin{center}
\leavevmode
\epsfxsize=0.45\textwidth
\epsffile{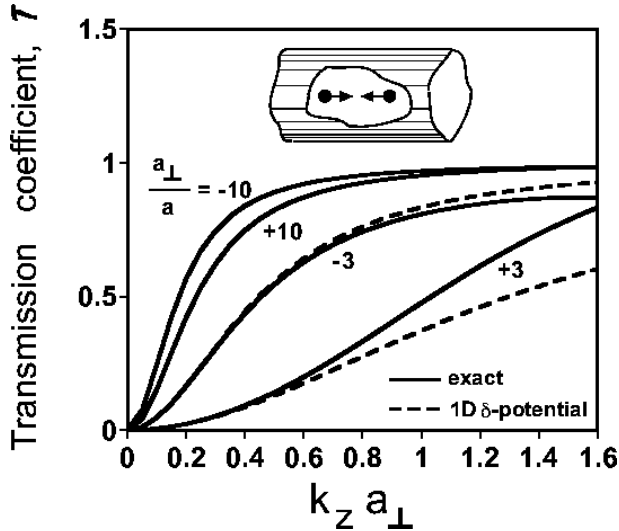}
\end{center}
\caption
{
Transmission probability as a function of the incident momentum.
Solid lines  correspond to the exact 1D scattering amplitude
(\ref{C_Amplitude}). Dashed lines correspond to the 1D $\delta$-potential
approximation $f_{\rm even} \approx f_{\rm even}^{\delta}$.
For $a_{\perp}/a = \pm 10$
the result of the $\delta$-potential approximation almost coincides
with the exact amplitude and is not shown.
\bigskip
\label{f_Transmission}
}
\end{figure}
For low velocities the exact scattering 
amplitude (\ref{C_Amplitude}) can be approximated by a   
scattering amplitude 
$f_{\rm even}^{\delta}(k_{z}) = - 1/(1 + i k_{z} a_{\rm 1D})$
for a {\it one-dimensional $\delta$-potential}
\begin{eqnarray}
U_{\rm 1D}(z) = g_{\rm 1D} \delta(z)
\label{1D_Potential}
\end{eqnarray}
of a coupling strength
\begin{eqnarray}
g_{\rm 1D} = -\, \frac{\hbar^2}{\mu \, a_{\rm 1D}} =
g \, |\phi_{0,0}(0)|^2 \, 
\left( 
 1-{\cal C}\frac{a}{a_{\perp}}
\right)^{-1} \, ,
\label{g_1D}
\end{eqnarray}
where $|\phi_{0,0}(0)|^2 =
1/\pi a_{\perp}^2$.
To illustrate this statement we plot (Fig.\ref{f_Transmission}) 
the transmission 
coefficient ${\cal T} = |1+f_{\rm even}+f_{\rm odd}|^2$ calculated using the 
exact scattering amplitude (\ref{C_Amplitude}) along with 
the results of the 1D $\delta$-potential approximation 
$f_{\rm even} \approx f_{\rm even}^{\delta}$.
Recall that in both cases the odd-wave scattering 
amplitude vanishes: $f_{\rm odd} = f_{\rm odd}^{\delta} = 0$.

%
%

The effective potential 
(\ref{1D_Potential},\ref{g_1D}) can be shown to 
reproduce the low-energy scattering properties of the radius $a$ hard spheres 
in presence of a transverse trap.
Recall that the Huang's potential (\ref{Huang}) plays exactly the same role in the 
free space.  
By analogy with the ``free-space'' case \cite{Huang}, the the limits 
of applicability of our potential  
can be extended to the {\it many-body problems} 
to describe interactions between the atoms   
confined   
to the ground transverse state. 

Note that the one-dimensional  gas of bosons interacting 
via a $\delta$-potential 
is already widely investigated in theoretical physics \cite{Int_Systems}.
One of the most interesting models belonging 
to this class is a {\it one-dimensional gas of impenetrable bosons}. 
Formally this 
model corresponds to the infinitely strong repulsive interaction 
between atoms: $g_{\rm 1D} \rightarrow +\infty$. More 
rigorously the impenetrable bosons regime corresponds to the 
low-energy scattering limit 
\begin{eqnarray}
k_{z} |a_{\rm 1D}| \ll 1 
\label{Low_Energy}
\end{eqnarray}
when the corresponding transmission coefficient ${\cal T}$
approaches zero (see Fig.\ref{f_Transmission}). (Recall that for positive interaction strength 
$g_{\rm 1D}$ the corresponding one-dimensional scattering length $a_{\rm 1D}$ is negative.)
A remarkable feature of the gas of impenetrable bosons 
is a possibility of a one-to-one mapping between the bosonic system and a 
gas of noninteracting fermions. 

\begin{figure}
\begin{center}
\leavevmode
\epsfxsize=0.45\textwidth
\epsffile{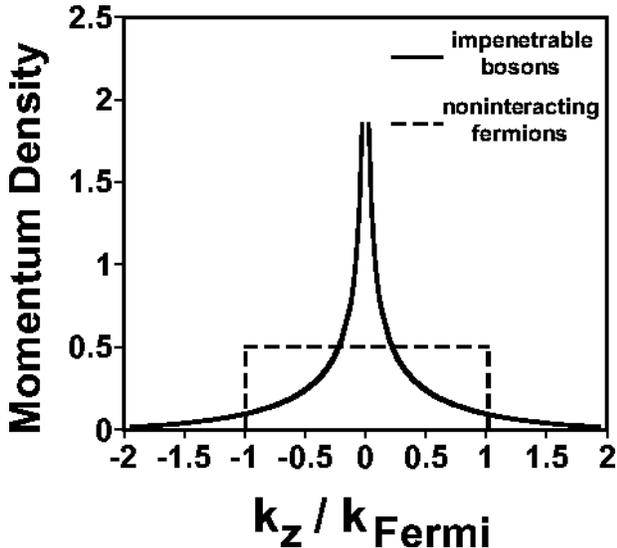}
\end{center}
\caption
{
Momentum distribution for a system of impenetrable bosons
at zero temperature in the
thermodynamic limit.
Corresponding distribution for an ideal Fermi gas
is shown for comparison.
\bigskip
\label{f_Momentum_Distribution}
}
\end{figure}

For $N$ impenetrable bosons confined in a periodic-boundary-conditions 
one-dimensional box of a length $L$ the ground state 
of the system $\Psi^{b}$ is given by an absolute value of the 
ground state of the $N$-particle ideal Fermi gas:
\begin{eqnarray}
\Psi^{b} = \left| \Psi^{f} \right| \,\, ,
\end{eqnarray}
where
\begin{eqnarray}
\Psi^{f}(z_1, z_2, \ldots, z_N) =
\frac{1}{\sqrt{N! L^{N}}}
 \mbox{det}(e^{i k_{j} z_{j^{\prime}}})
\\
k_{j} =
\frac{2\pi}{L} j \in \lbrack - k_{\rm Fermi}; \, + k_{\rm Fermi} \rbrack
\end{eqnarray}
and $k_{\rm Fermi} = \pi (N-1)/L$ is the one-dimensional Fermi radius. 
In
Fig.\ref{f_Momentum_Distribution} the zero-temperature
one-body momentum distribution $w(k_{z})$ for a system of 
impenetrable bosons in the
thermodynamic limit is shown. (It is normalized as 
$\int_{-\infty}^{+\infty} (dk_{z}/k_{\rm Fermi}) \, w(k_{z}) = 1$.)
The distribution has been calculated using the short-range and
long-range expansions for the one-body spatial correlation function 
given in
\cite{Vaidya}. 
The shape of the peak at the origin is given by 
$w(k_{z}) \approx \rho_{\infty}\sqrt{k_{\rm Fermi}/2\pi k_z}$ 
where 
$\rho_{\infty} = \pi e^{1/2} 2^{-1/3} A^{-6} = 0.92418\ldots$,
and $A = 1.2824\ldots$ is the Glaisher's constant.
Note that for finite values of $N$ and $L$  
the thermodynamic limit fails at 
small momenta
$k_{z} \sim 2\pi/L \sim 2 k_{\rm Fermi}/N$ 
and the momentum density curve is not valid in this domain. 
For comparison, we
plot also the corresponding momentum distribution for an ideal Fermi gas.  

The system of impenetrable bosons may be realized in the 
atom wave guides \cite{Waveguides} combined with a longitudinal 
confinement between two potential barriers. 
The impenetrable bosons regime 
may be identified by a presence of the $1/\sqrt{k_{z}}$ peak  
in the 
momentum distribution of atoms (see Fig.\ref{f_Momentum_Distribution}).  
Note that the low energy
scattering condition (\ref{Low_Energy}) puts an upper bound 
to the number of atoms. 
Indeed, using the fact that the maximal momentum of the relative motion 
between a pair of atoms
is given by 
$k_{max} = (\mu/m) \max_{j, j^{\prime}}(k_{j} - k_{j^{\prime}}) 
= k_{\rm Fermi}$  
one could show that
the low energy
scattering condition $k_{max} |a_{\rm 1D}| \ll 1$ leads 
to a limit for the number of atoms 
\begin{eqnarray}
N \ll N^* = \frac{L}{\pi |a_{\rm 1D}|} 
\propto 
L\, \omega_{\perp} \,\, ,
\end{eqnarray}
where $a_{\rm 1D}$ is given by (\ref{a_1D}). For
$\omega_{\perp} = 2\pi \times 10^4 \, {\rm Hz}$ and $L = 3 \, {\rm cm}$ 
the upper bound for the 
number of atoms is given by: 
$N^*_{\rm Rb} = 5 \times 10^3$ for rubidium 87 ($F=2, m_F=+2,
a =  +110 \, a_{\rm Bohr}$ \cite{Rb_aScatt});
and  
$N^*_{\rm Na} = 6 \times 10^2$ 
for sodium ($F=1, m_F=-1,
a = +52 \, a_{\rm Bohr}$ \cite{Na_aScatt}).
Note that the above numbers can be improved in 
stronger traps since the bound $N^*$ increases approximately linearly 
with the confinement frequency. 

In conclusion, we calculate the one-dimensional 
scattering amplitude and effective
one-dimensional interaction potential for atoms transversally confined 
by a two-dimensional harmonic potential.
We suggest 
a realization of the Tonks gas -- 
a one-dimensional
gas of impenetrable
bosons. 
From the experimental point of view the impenetrable
bosons regime corresponds to highly elongated traps (waveguides) at 
low temperatures ($k_{\rm B}T \ll \hbar \omega_{\perp}$) 
and at low linear densities (\,$\rho a \ll (2\pi)^{-1}(a/a_{\perp})^2$, $a>0$). 
We give an estimate for suitable experimental
parameters for
alkali atoms confined in waveguides.

We would like 
to express our appreciation for many useful 
discussions with Y. Castin, R. Dum, K. Johnson, M. Prentiss, E. Heller, 
A. Lupu-Sax,
M. Naraschewski, V. Lorent, S. P. Smith, C. Herzog, C. A. Tracy, V. E. Korepin, and 
J. M. Doyle.


M.O. was supported by the National Science Foundation
grant for light force dynamics \#PHY-93-12572 and by the grant {\it PAST}
of the French Government.
This work was also partially supported by
the NSF through
a grant for the Institute for Theoretical Atomic and Molecular
Physics at Harvard University and the Smithsonian Astrophysical 
Observatory.
Laboratoire Kastler-Brossel is an {\it unit\'{e} de recherche de 
l'Ecole Normale Sup\'{e}rieure et de l'Universit\'{e} Pierre et Marie
Curie, associ\'{e}e au CNRS}.
%
%

%


\begin{thebibliography}{99}
\bibitem{Epsilon}
Recall that according to the condition (\ref{C_Kinetic_Constr}) 
the value of $\epsilon$ is bounded from below as $\epsilon > -1$.
\end{thebibliography}
\end{document}